\documentclass[10pt,twocolumn,letterpaper]{article}

\usepackage{cvpr_teaser}
\usepackage{times}
\usepackage{epsfig}
\usepackage{graphicx}
\usepackage{amsmath}
\usepackage{amssymb}
\usepackage{color}
\usepackage{multirow}
\usepackage{booktabs}
\usepackage{paralist}
\usepackage{mathrsfs}
\usepackage{bbm}
\usepackage{epstopdf}
\usepackage[pagebackref=true,breaklinks=true,letterpaper=true,colorlinks,linkcolor=blue,citecolor=blue,bookmarks=false]{hyperref}

\cvprfinalcopy %

\makeatletter
\newcommand{\rmnum}[1]{\romannumeral #1}
\newcommand{\Rmnum}[1]{\expandafter\@slowromancap\romannumeral #1@}
\makeatother

\def\cvprPaperID{921} %

\ifcvprfinal\fi
\begin{document}

\title{Towards Real Scene Super-Resolution with Raw Images}

\author{Xiangyu Xu \quad Yongrui Ma \quad Wenxiu Sun\\
	SenseTime Research
}

\maketitle
\begin{figure}[!t]\footnotesize
	\vspace{-3.5in}
	\begin{minipage}{\textwidth}
		\begin{center}
			\newcommand{\figwidth}{0.035\linewidth}
			\newcommand{\Figwidth}{0.135\linewidth}
			\newcommand{\shiftfigure}{\hspace{-3.0mm}}
			\begin{tabular}{c}
				\includegraphics[width = 0.90\linewidth]{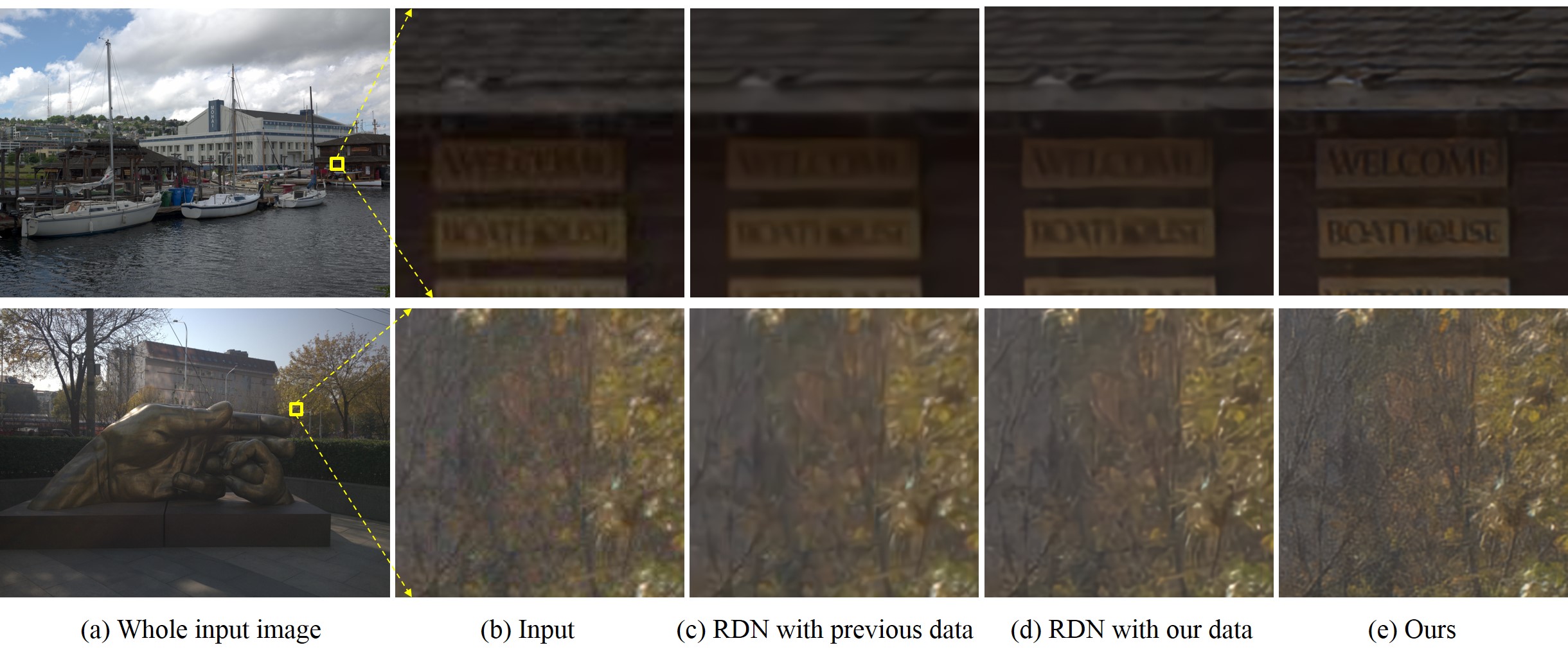} \\
			\end{tabular}
		\end{center}
		\vspace{-2mm}
		\caption{%
			Existing super-resolution method (RDN~\cite{SR-residual-dense}) does not perform well for real captured images as shown in (c).
			We can obtain sharper results (d) by re-training existing model~\cite{SR-residual-dense} with the data generated by our method.
			Furthermore, we recover more structures and details (e) by exploiting the radiance information recorded in raw images.
			The two input images (a) are captured by Leica SL Typ-601
			 and iPhone 6s respectively, and both cameras are not seen by the models during training.
			Results best viewed electronically with zoom.}
		\label{fig:teaser} \end{minipage}
	\vspace{-2mm}
\end{figure}

\begin{abstract}
Most existing super-resolution methods do not perform well in real scenarios due to lack of realistic training data and information loss of the model input.
To solve the first problem, we propose a new pipeline to generate realistic training data by simulating the imaging process of digital cameras.
And to remedy the information loss of the input, we develop a dual convolutional neural network to exploit the originally captured radiance information in raw images.
In addition, we propose to learn a spatially-variant color transformation which helps more effective color corrections.
Extensive experiments demonstrate that super-resolution with raw data helps recover fine details and clear structures, and more importantly, the proposed network and data generation pipeline achieve superior results for single image super-resolution in real scenarios.

\end{abstract}

\vspace{-4mm}

\section{Introduction}
\label{sec:introduction}
In optical photography, the number of pixels representing an object, \ie the image resolution, is directly proportional to square of the focal length of the camera~\cite{geo_optics}.
While one can use a long-focus lens to obtain a high-resolution image, the range of the captured scene is usually limited by the size of the sensor array at the image plane.
Thus, it is often desirable for users to capture the wide-range scene at a lower resolution with a short-focus camera (\eg a wide-angle lens), 
and then apply the single image super-resolution technique which recovers a high-resolution image from its low-resolution version.

Most state-of-the-art super-resolution methods~\cite{SRCNN,SR-VDSR,SR-regression-accv14,SR-SCN,SR-laplacian,SR-dense,SR-residual-dense,xxy-iccv17} are based on data-driven models, and in particular, deep convolutional neural networks (CNNs)~\cite{SRCNN,xxy-tip19-dehazing,xxy-icip18-temporal}.
While these methods are effective on synthetic data, they do not perform well for real captured images by cameras or cellphones (examples shown in Figure~\ref{fig:teaser}(c)) owing to lack of realistic training data and information loss of network input.
To remedy these problems and enable real scene super-resolution, we propose a new pipeline for generating training data and a dual CNN model for exploiting additional raw information, which are described as below.

First, most existing methods cannot synthesize realistic training data; the low-resolution images are usually generated with a fixed downsampling blur kernel (\eg bicubic kernel) and homoscedastic Gaussian noise~\cite{SR-residual-dense,xxy-tip18-edgedeblur}.
On one hand, the blur kernel in practice may vary with zoom, focus, and camera shake during image capturing, which is beyond the fixed kernel assumption.
On the other hand, image noise usually obeys heteroscedastic Gaussian distribution~\cite{benchmarking_denoising} whose variance depends on the pixel intensity, which is in sharp contrast to the homoscedastic Gaussian noise.
More importantly, both the blur kernel and noise should be applied to the linear raw data whereas previous approaches use the pre-processed non-linear color images.
To solve above problems, 
we synthesize the training data in linear space by simulating the imaging process of digital cameras, and applying different kernels and heteroscedastic Gaussian noise to approximate real scenarios.
As shown in Figure~\ref{fig:teaser}(d), we can obtain sharper results by training existing model~\cite{SR-residual-dense} with the data from our generation pipeline. 

\begin{figure}[t]
	\footnotesize
	\begin{center}
		\begin{tabular}{c}
			\includegraphics[width = 0.95\linewidth]{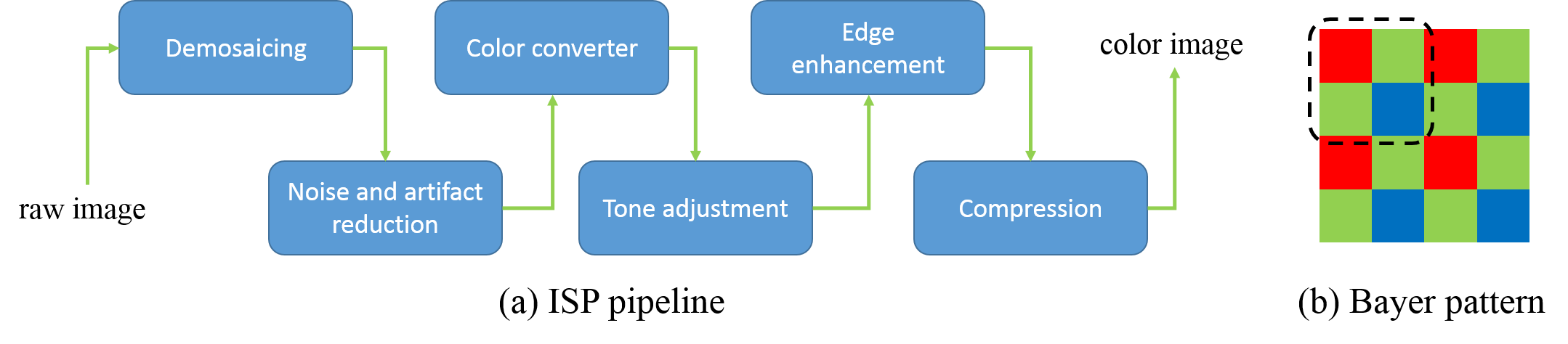} \\
		\end{tabular}
	\end{center}
	\caption{%
		Typical ISP pipeline of digital cameras.
		(b) represents a Bayer pattern raw image, where the sensels surrounded by the black-dot curve is a Bayer pattern block. 
		}
	\label{fig:introduction}
\end{figure}

Second,
while modern cameras provide both the raw data and the pre-processed color image (produced by the image signal processing system, \ie ISP) to users~\cite{hdr+},
most super-resolution algorithms only take the color image as input, which does not make full use of the radiance information existing in raw data.
By contrast, we directly use raw data for restoring high-resolution clear images, which conveys several advantages:
(\rmnum{1}) More information could be exploited in raw pixels since they are typically 12 or 14 bits~\cite{adobe-mit-5k}, %
whereas the color pixels produced by ISP are typically 8 bits~\cite{hdr+}. 
We show a typical ISP pipeline in Figure~\ref{fig:introduction}(a).
Except for the bit depth, there is additional information loss within the ISP pipeline, such as the noise reduction and compression~\cite{pennebaker1992jpeg}.
(\rmnum{2}) 
Raw data is proportional to scene radiance, while the ISP contains nonlinear operations, such as tone mapping.
Thus, the linear degradations in the imaging process, including blur and noise, are nonlinear in the processed RGB space, which brings more difficulties in image restoration~\cite{raw-reconstruction}.
(\rmnum{3}) %
The demosaicing step in the ISP is highly related to super-resolution, because these two problems both refer to the resolution limitations of cameras~\cite{SR_demosaic_deep}. 
Therefore, to solve the super-resolution problem with pre-processed images is sub-optimal and could be inferior to a single unified model solving both problems simultaneously.

\begin{figure*}[t]
	\footnotesize
	\begin{center}
		\begin{tabular}{c}
			\includegraphics[width = 0.95\linewidth]{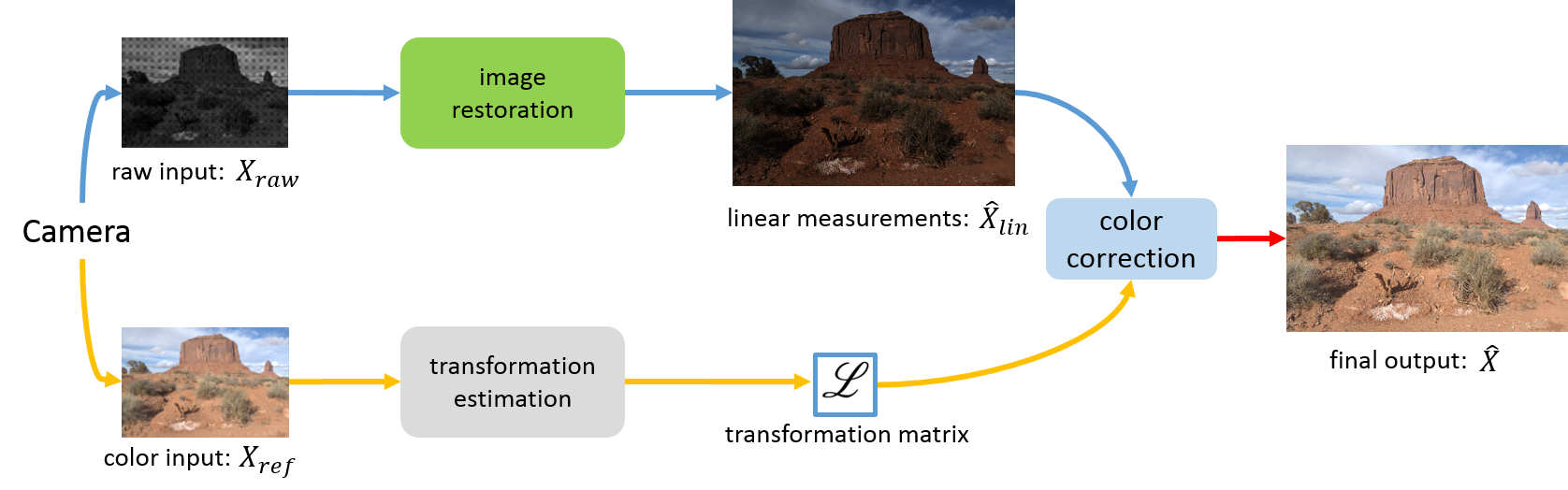} \\
		\end{tabular}
	\end{center}
	\caption{Overview of the proposed network. 
		Our model has two parallel branches,
		where the first branch exploits raw data $X_{raw}$ to restore high-resolution linear measurements $\widehat{X}_{lin}$ for all color channels with clear structures and fine details,
		and the second branch estimates the transformation matrix to recover the final color result $\widehat{X}$ using the low-resolution color image $X_{ref}$ as reference.
	}
	\label{fig:network_overview}
\end{figure*}
In this paper, we introduce a new super-resolution method to exploit raw data from camera sensors.
Existing raw image processing networks~\cite{learning_to_see_in_the_dark,deepisp_tip} usually learn a direct mapping function from the degraded raw image to the desired full color output.
However, raw data does not have the relevant information for color corrections conducted within the ISP system, and thereby the networks trained with it could only be used for one specific camera.
To solve this problem,
we propose a dual CNN architecture (Figure~\ref{fig:network_overview}) which takes both the degraded raw and color images as input, so that our model could generalize well to different cameras.
The proposed model consists of two parallel branches,
where the first branch restores clear structures and fine details with the raw data, 
and the second branch recovers high-fidelity colors with the low-resolution RGB image as reference.
To exploit multi-scale features, we use densely connected convolution layers~\cite{densenet} in an encoder-decoder framework for image restoration.
For the color correction branch, simply adopting the technique in \cite{deepisp_tip} to learn a global transformation usually leads to artifacts and incorrect color appearances.
To address this issue, we propose to learn pixel-wise color transformations to handle more complex spatially-variant color operations and generate more appealing results. 
In addition, we introduce feature fusion for more accurate color correction estimation.
As shown in Figure~\ref{fig:teaser}(e), the proposed algorithm significantly improves the super-resolution results for real captured images.

The main contributions of this work are summarized as follows. 
First, we design a new data generation pipeline which enables synthesizing realistic raw and color training data for image super-resolution.
Second, we develop a dual network architecture to exploit both raw data and color images for real scene super-resolution, which is able to generalize to different cameras.
In addition, we propose to learn spatial-variant color transformations as well as feature fusion for better performance.
Extensive experiments demonstrate that solving the problem using raw data helps recover fine details and clear structures, and more importantly, the proposed network and data generation pipeline achieve superior results for single image super-resolution in real scenarios.

\section{Related Work}
We discuss state-of-the-art super-resolution methods as well as learning-based raw image processing, and put this work in proper context.

\vspace{1ex}
{\noindent \bf{Super-resolution.}} %
Most state-of-the-art super-resolution methods~\cite{SR-SCN,SR-VDSR,SR-dense,SRCNN,SR-residual-dense,SR-laplacian,FSRCNN,xxy-iccv17} learn CNNs to reconstruct high-resolution images from low-resolution color inputs.
Dong \etal~\cite{SRCNN} propose a three-layer CNN for mapping the low-resolution patches to high-resolution space, but fail to get better results with deeper networks~\cite{SRCNN_pami}.
To solve this problem, Kim \etal~\cite{SR-VDSR} introduce residual learning to accelerate training and achieve better results.
Tong \etal~\cite{SR-dense} use dense skip connections to further speed up the reconstruction process.
While these methods are effective in interpolating pixels, they are based on preprocessed color images and thus have limitations in producing realistic details. 
By contrast, we propose to exploit both raw data and color image in a unified framework for better super-resolution.

\vspace{1ex}
{\noindent \bf{Joint super-resolution and demosaicing.}} %
Many existing methods for this problem estimate a high-resolution color image with multiple low-resolution frames~\cite{farsiu2006multiframe,vandewalle2007joint}. 
More closely related to our task, Zhou \etal~\cite{SR_demosaic_deep} propose a deep residual network for single image super-resolution with mosaiced images.
However, this model is trained on gamma-corrected image pairs which may not work well for real linear data.
More importantly, these works do not consider the complex color correction steps applied by camera ISPs, and thus cannot recover high-fidelity color appearances.
Different from them, the proposed algorithm solves the problems of image restoration and color correction simultaneously, which are more suitable for real applications.

\vspace{1ex}
{\noindent \bf{Learning-based raw image processing.}} %
In recent years, learning-based methods have been proposed for raw image processing~\cite{jiang2017learning,learning_to_see_in_the_dark,deepisp_tip}.
Jiang \etal~\cite{jiang2017learning} propose to learn a large collection of local linear filters to approximate the complex nonlinear ISP pipelines.
Following their work, Schwartz \etal~\cite{deepisp_tip} use deep CNNs for learning the color correction operations of specific digital cameras.
Chen~\etal~\cite{learning_to_see_in_the_dark} train a neural network with raw data as input for fast low-light imaging.
In this work, we learn color correction in the context of raw image super-resolution. Instead of learning a color correction pipeline for one specific camera, we use a low-resolution color image as reference for handling images from more diverse ISP systems.

\begin{figure*}[t]
	\footnotesize
	\begin{center}
		\begin{tabular}{c}
			\includegraphics[width = 0.95\linewidth]{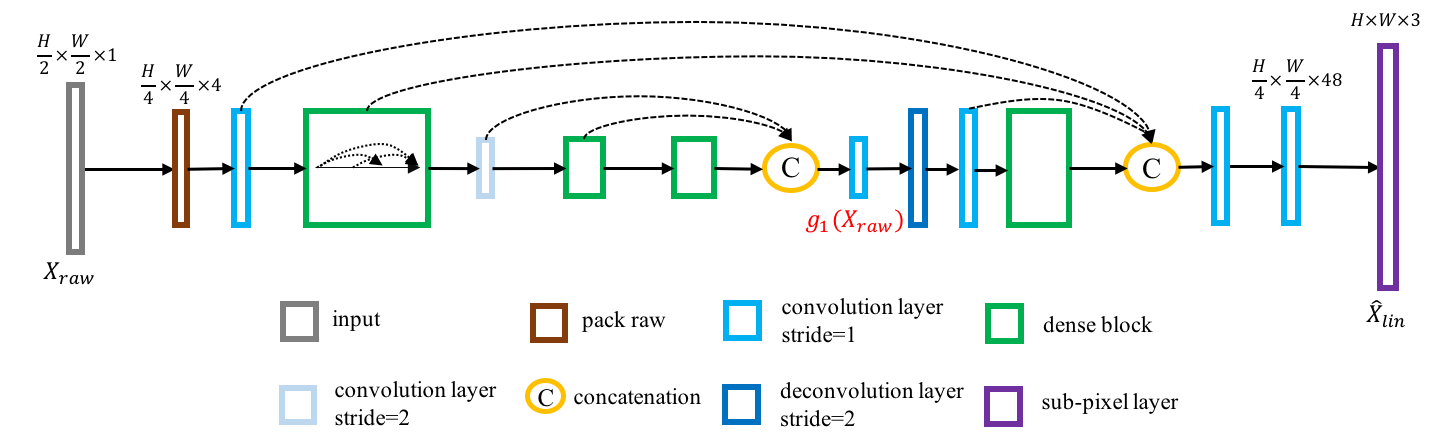} \\
		\end{tabular}
	\end{center}
	\caption{%
		The image restoration branch adopts dense blocks in an encoder-decoder framework,
		which reconstructs high-resolution linear color measurements $\widehat{X}_{lin}$ from the degraded low-resolution raw input $X_{raw}$. 
	}
	\label{fig:branch1}
\end{figure*}
\section{Method}
For better super-resolution results in real scenarios, we propose a new data generation pipeline to synthesize more realistic training data, and a dual CNN model to exploit the radiance information recorded in raw data.
We describe the synthetic data generation pipeline in Section~\ref{sec:training_data} and the network architecture in Section~\ref{sec:network}.

\subsection{Training Data}
\label{sec:training_data}
Most super-resolution methods~\cite{SR-VDSR,SRCNN} generate training data by downsampling high-resolution color images with a fixed bicubic blur kernel.
And homoscedastic Gaussian noise is often used to model real image noise~\cite{SR-residual-dense}.
However, as introduced in Section~\ref{sec:introduction}, the low-resolution images generated in this way does not resemble real captured images and will be less effective for training real scene super-resolution models.
Moreover, we need to generate low-resolution raw data as well for training the proposed dual CNN, which is often approached by directly mosaicing the low-resolution color images~\cite{demosaicking_denoising_deep,SR_demosaic_deep}.
This strategy ignores the fact that the color images have been processed by nonlinear operations of the ISP system  while the raw data should be from linear color measurements of the pixels.
To solve these problems, we start with high-quality raw images~\cite{adobe-mit-5k} and generate realistic training data by simulating the imaging process of the degraded images.
We first synthesize ground truth linear color measurements by downsampling the high-quality raw images so that each pixel could have its ground truth red, green and blue values.
In particular, for Bayer pattern raw data (Figure~\ref{fig:introduction}(b)), we define a block of Bayer pattern sensels as one new virtual sensel, 
where all color measurements are available.
In this way, we can obtain the desired images $X_{lin} \in \mathbb{R}^{H\times W \times 3}$ with linear measurements of all three colors for each pixel.
$H$ and $W$ denote the height and width of the ground truth linear image.
Similar with~\cite{demosaic_denoise_randomfields}, we compensate the color shift artifact by aligning the center of mass of each color in the new sensel.
With the ground truth linear measurements $X_{lin}$, we can easily generate the ground truth color images $X_{gt} \in \mathbb{R}^{H\times W \times 3}$ by simulating the color correction steps of the camera ISP system, such as color space conversion and tone adjustment. 
For the simulation, we use Dcraw~\cite{dcraw}  which is a widely-used raw processing software.

To generate degraded low-resolution raw images $X_{raw} \in \mathbb{R}^{\frac{H}{2}\times \frac{W}{2}}$, we separately apply blurring, downsampling, Bayer sampling, and noise onto the linear color measurements:
\begin{align}
X_{raw}=f_{Bayer}(f_{down}(X_{lin}*k_{def}*k_{mot}))+n,
\end{align}
where $f_{Bayer}$ is the Bayer sampling function which mosaics images in accordance with the Bayer pattern (Figure~\ref{fig:introduction}(b)).
$f_{down}$ represents the downsampling function with a sampling factor of two.
Since the imaging process is likely to be affected by out-of-focus effect as well as camera shake,
we consider both defocus blur $k_{def}$ modeled as disk kernels with variant sizes, and modest motion blur $k_{mot}$ generated by random walk~\cite{schuler2013machine}.
$*$ denotes the convolution operator. 
To synthesize more realistic training data,
we add heteroscedastic Gaussian noise~\cite{KPN,benchmarking_denoising,healey1994radiometric} to the generated raw data:
\begin{align}
n(x) \sim \mathcal{N}(0, \sigma_1^2 x+\sigma_2^2),
\end{align}
where the variance of noise $n$ depends on the pixel intensity $x$. $\sigma_1$ and $\sigma_2$ are parameters of the noise.
Finally,  the raw image $X_{raw}$ is demosaiced with the AHD~\cite{AHD2005}, a commonly-used demosaicing method, and further processed by Dcraw to produce the low-resolution color image $X_{ref} \in \mathbb{R}^{\frac{H}{2}\times \frac{W}{2} \times 3}$.  
We compress $X_{ref}$ as 8-bit JPEG as normally done in digital cameras.
Note that the settings for the color correction steps in Dcraw are the same as those used in generating $X_{gt}$, so that the reference colors correspond well to the ground truth.
The proposed pipeline synthesizes realistic data so that the models trained on $X_{raw}$ and $X_{ref}$ generalize well to real degraded raw and color images (Figure~\ref{fig:teaser}(e)).

\subsection{Network Architecture}
\label{sec:network}
A straightforward way to exploit raw data for super-resolution is to directly learn a mapping function from raw images to high-resolution color images with neural networks.
While the raw data is advantageous for image restoration,
this naive strategy does not work well in practice;
because the raw data does not have the relevant information for color correction and tone enhancement which have been conducted within the ISP system of digital cameras.
In fact, one raw image could potentially correspond to multiple ground truth color images generated by different image processing algorithms of various ISP pipelines, which will confuse the network training and make it non-practical to train a model to reconstruct high-resolution images with desired colors.
To solve this problem, we propose a two-branch CNN as shown in Figure~\ref{fig:network_overview}, where the first branch exploits raw data to restore high-resolution linear measurements for all color channels with clear structures and fine details, 
and the second branch estimates the transformation matrix to recover high-fidelity color appearances using the low-resolution color image as reference.
The reference image complements the limitations of raw data; thus, jointly training two branches with raw and color data could essentially help recover better results.

\vspace{1ex}
{{\noindent \bf{Image restoration.}}} %
We show an illustration of the image restoration network in Figure~\ref{fig:branch1}.
Similar with~\cite{learning_to_see_in_the_dark}, we first pack the raw data $X_{raw}$ into four channels which respectively correspond to the R, G, B, G patterns in Figure~\ref{fig:introduction}(b). %
Then we apply a convolution layer to the packed four-channel input for learning low-level features, and consecutively adopt four dense blocks~\cite{densenet,SR-dense} to learn more complex nonlinear mappings for image restoration.
The dense block is composed of eight densely connected convolution layers.
Different from previous work~\cite{densenet,SR-dense}, we deploy the dense blocks in an encoder-decoder framework to make the model more compact and efficient. We use $g_1(X_{raw})$ to denote the encoded features as shown in Figure~\ref{fig:branch1}.
Finally, the feature maps with the same spatial size are concatenated and fed into a convolution layer, and this layer produces $48$ feature maps which are rearranged by a sub-pixel layer~\cite{shi2016real} to restore the three-channel linear measurements  $\widehat{X}_{lin} \in \mathbb{R}^{H\times W \times 3}$  with $2$ times resolution of the input.

\begin{figure}[t]
	\footnotesize
	\begin{center}
		\begin{tabular}{c}
			\includegraphics[width = 1\linewidth]{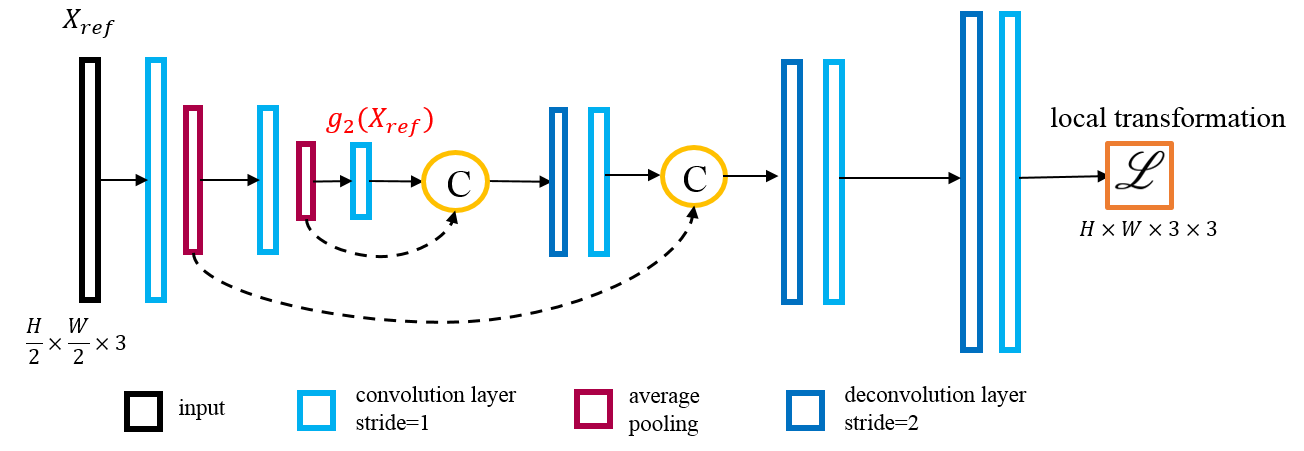} \\
		\end{tabular}
	\end{center}
	\caption{%
		Network architecture of the color correction branch.
		Our model predicts the pixel-wise transformation $\mathscr{L}(X_{ref})$ with a reference color image.
	}
	\label{fig:branch2}
\end{figure}

\vspace{1ex}
{\noindent \bf{Color correction.}} 
With the predicted linear measurements  $\widehat{X}_{lin}$, we learn the second branch for color correction which reconstructs the final result $\widehat{X}  \in \mathbb{R}^{H\times W \times 3}$ using the color reference image $X_{ref}$.
Similar with~\cite{deepisp_tip}, 
we use a CNN to estimate a transformation $\mathscr{G}(X_{ref}) \in \mathbb{R}^{3\times 3}$ to produce a global color correction of the image.
Mathematically, the correction could be formulated as:  $\mathscr{G}(X_{ref}) \widehat{X}_{lin}[i, j]$,
where $\widehat{X}_{lin}[i,j] \in \mathbb{R}^{3\times 1}$ represents the RGB values at pixel $[i,j]$ of the linear image.
However, this global correction strategy does not work well when the color transformation of camera ISP involves spatially-variant operations. 

To solve this problem, 
we propose to learn a pixel-wise transformation $\mathscr{L}(X_{ref}) \in \mathbb{R}^{H\times W \times 3\times 3}$ which allows different color corrections for each spatial location. 
Thus, we can generate the final results as:
\begin{align}
\widehat{X}[i,j]=\mathscr{L}(X_{ref})[i,j] \widehat{X}_{lin}[i, j],
\end{align}
where $\mathscr{L}(X_{ref})[i,j] \in \mathbb{R}^{3\times 3}$ is the local transformation matrix for the RGB vector at pixel $[i,j]$. 
Note that we directly transform the RGB vectors instead of using the quadratic form in~\cite{deepisp_tip} as we empirically find no benefits from it.

We adopt a U-Net structure~\cite{xxy-eccv18-monodepth,xxy-eccv18-rendering} for predicting the transformation $\mathscr{L}(X_{ref})$ in Figure~\ref{fig:branch2}.
The CNN starts with an encoder including three convolution layers with $2\times2$ average pooling to extract the encoded features $g_2(X_{ref})$.
To estimate the spatially-variant transformations, 
we adopt a decoder consisting of consecutive deconvolution and convolution layers, which expands the encoded features to the desired resolution $H\times W$.
We concatenate the feature maps with the same resolution in the encoder and decoder to exploit hierarchy features.
Finally, the output layer of the decoder  generates $3\times 3$ weights for each pixel, which form the pixel-wise transformation matrix $\mathscr{L}(X_{ref})$.

\vspace{1ex}
{\noindent \bf{Feature fusion.}} 
As the transformation $\mathscr{L}(X_{ref})$ is applied to the restored image,
it would be beneficial to make the matrix estimation network aware of the features in the image restoration branch, so that $\mathscr{L}(X_{ref})$ could more accurately adapt to the structures of the restored image.
Towards this end,
we develop feature fusion for $g_2(X_{ref})$ using the encoded features from the first branch $g_1(X_{raw})$.
As $g_2(X_{ref})$ and $g_1(X_{raw})$ are likely to have different scales, we fuse them by weighted sum where the weights are learned to adaptively update the features. 
We could formulate the feature fusion as:
\begin{align}
	g_2(X_{ref}) \leftarrow g_2(X_{ref})+\phi (g_1(X_{raw})),
\end{align}
where $\phi$ is a $1 \times 1$ convolution. 
After this, we put the updated $g_2(X_{ref})$ back into the original branch, and the following decoder could further use the fused features for transformation estimation.
The convolutions are initialized with zeros to avoid interrupting the initial behavior of the color correction branch.
\begin{figure*}[t]
	\footnotesize
	\begin{center}
		\begin{tabular}{c}
			\includegraphics[width = 0.99\linewidth]{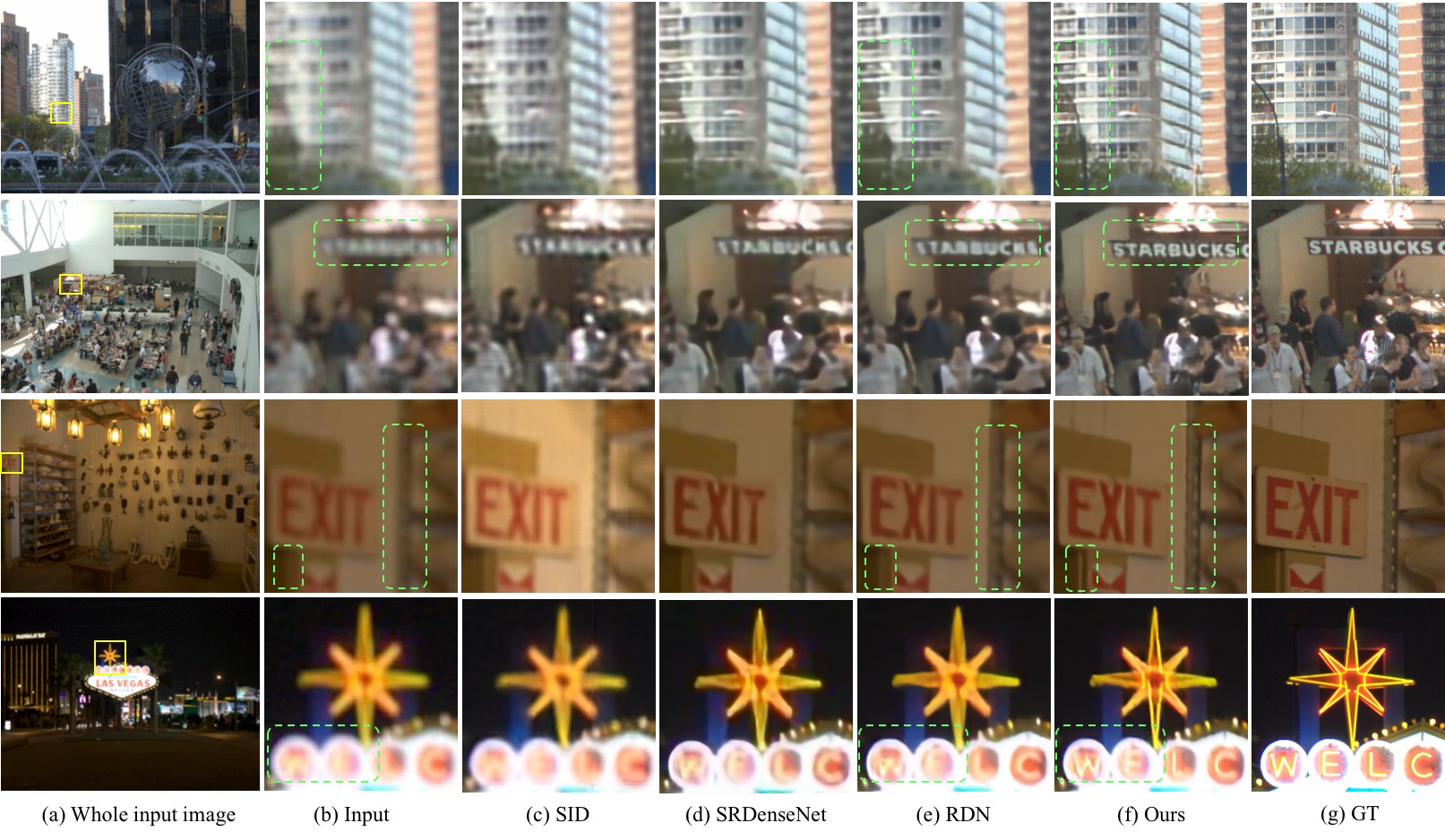} \\
		\end{tabular}
	\end{center}
	\caption{%
		Results on our synthetic dataset. References for the baseline methods could be found in Table~\ref{tab:syn1}. ``GT" represent ground truth.
	}
	\label{fig:synthetic}
\end{figure*}

\section{Experimental Results}
We describe the implementation details of our method and present analysis and evaluations on both synthetic and real data in this section. 
\subsection{Implementation Details}
For generating training data, we use the MIT-Adobe 5K dataset~\cite{adobe-mit-5k} which is composed of $5000$ raw photographs with size around $2000\times 3000$.
After manually removing images with noticeable noise and blur,
we obtain $1300$ high-quality images for training and $150$ for testing which are captured by $19$ types of Canon cameras.
We use the method described in Section~\ref{sec:training_data} to synthesize the training and test datasets.
The radius of the defocus blur is randomly sampled from $[1, 5]$ and the maximum size of the motion kernel is sampled from $[3, 11]$.
The noise parameters $\sigma_1, \sigma_2$ are respectively sampled from $[0,10^{-2}]$ and $[0,10^{-3}]$.

We use the Xavier initializer~\cite{glorot2010understanding} for the network weights and the LeakyReLU~\cite{leakyRelu} with slope 0.2 as the activation function.
We adopt the $L_1$ loss function for training the network and use the Adam optimizer~\cite{kingma2014adam} with initial learning rate $2\times 10^{-4}$.
We crop $256\times 256$ patches as input and use a batch size of $6$. 
We first train the model with $4\times10^{4}$ iterations where the learning rate is decreased by a factor of $0.96$ every $2\times 10^3$ updates, and then another $4\times10^{4}$ iterations at a lower rate $10^{-5}$.
During the test phase, 
we chop the input into overlapping patches and process each patch separately.
The reconstructed high-resolution patches are placed back to the corresponding locations and averaged in overlapping regions.

\vspace{1ex}
{\noindent \bf{Baseline methods.}} %
We compare our method with state-of-the-art super-resolution methods~\cite{SRCNN,SR-VDSR,SR-dense,SR-residual-dense} as well as learning-based raw image processing algorithms~\cite{learning_to_see_in_the_dark,deepisp_tip}.
As the raw image processing networks~\cite{learning_to_see_in_the_dark,deepisp_tip} are not originally designed for super-resolution, we add deconvolution layers after them for increasing the output resolution.
All the baseline models are trained on our data with the same settings.

\begin{figure*}[t]
	\footnotesize
	\begin{center}
		\begin{tabular}{c}
			\includegraphics[width = 0.99\linewidth]{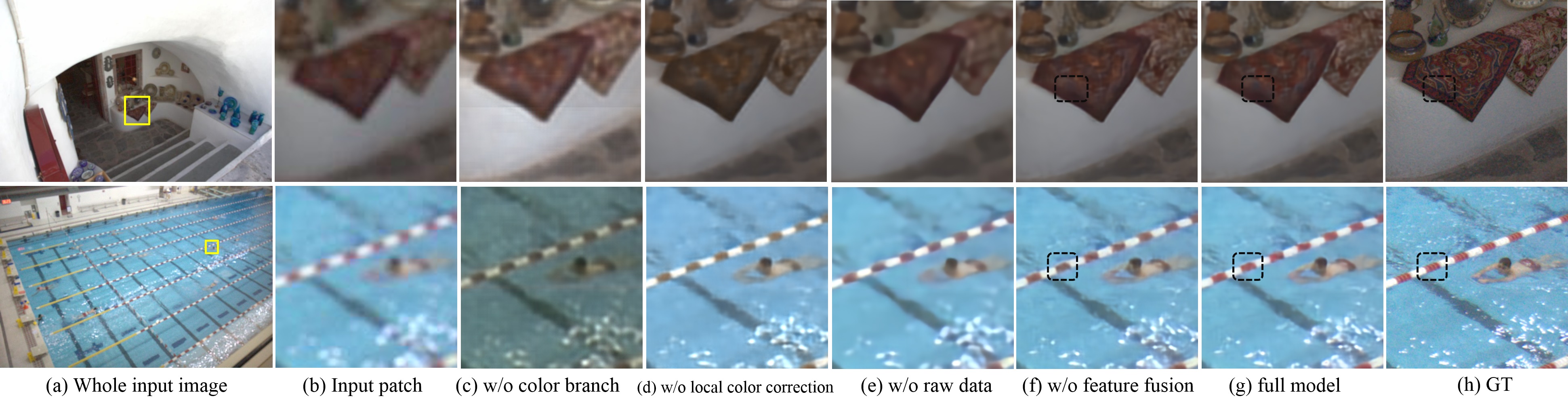} \\
		\end{tabular}
	\end{center}
	\caption{%
		Ablation study of the proposed network on the synthetic dataset. 
	}
	\label{fig:ablation}
\end{figure*}
\begin{figure*}[t]
	\footnotesize
	\begin{center}
		\begin{tabular}{c}
			\includegraphics[width = 0.99\linewidth]{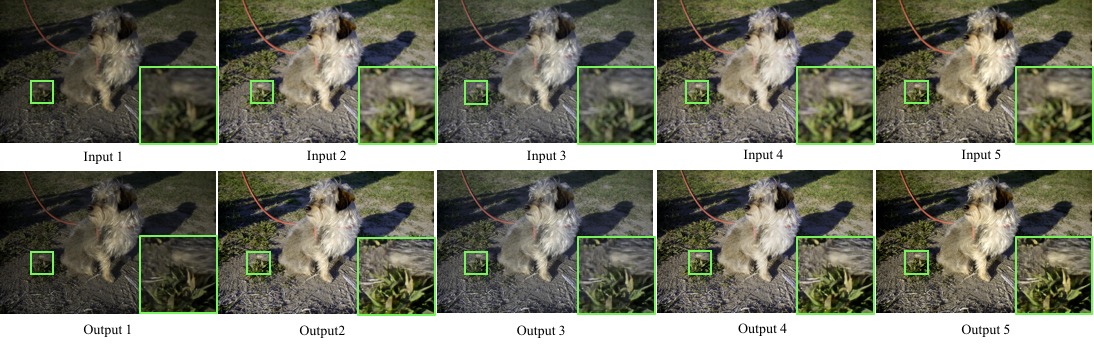} \\
		\end{tabular}
	\end{center}
	\caption{%
		Generalization to manually retouched color images from MIT-Adobe 5K dataset~\cite{adobe-mit-5k}. The proposed method generates clear structures and high-fidelity colors.
	}
	\label{fig:artistic}
\end{figure*}

\begin{table}[t]
	\caption{\label{tab:syn1}Quantitative evaluations on the proposed synthetic dataset. ``Blind" represents the images with variable blur kernels, and ``Non-blind" denotes fixed kernel.
		{\color{red}Red} and {\color{blue}blue} text indicates the first and second best performance, respectively.
	}
	\footnotesize
	\centering
	\begin{tabular}{|c|c|c|c|c|}
		\hline
		\multicolumn{1}{|c|}{\multirow{2}*{Methods}}&\multicolumn{2}{|c|}{Blind}&\multicolumn{2}{|c|}{Non-blind}\\
		\cline{2-5}
		& PSNR & SSIM & PSNR & SSIM \\    %
		\hline
		SID~\cite{learning_to_see_in_the_dark} & 21.87 & 0.7325 & 21.89 & 0.7346  \\
		DeepISP~\cite{deepisp_tip}& 21.71 & 0.7323 & 21.84 & 0.7382  \\
		\hline
		SRCNN~\cite{SRCNN}& 28.83 & 0.7556 & 29.19 & 0.7625  \\
		VDSR~\cite{SR-VDSR} & 29.32 & 0.7686 & 29.88 & 0.7803  \\
		SRDenseNet~\cite{SR-dense} & 29.46 & 0.7732 & 30.05 & 0.7844  \\
		RDN~\cite{SR-residual-dense}& 29.93 & 0.7804 & 30.46 & 0.7897  \\
		\hline
		w/o color branch& 20.54 & 0.7252  & 21.13 & 0.7413 \\
		w/o local color correction& 29.81 & 0.7954 & 30.29 & 0.8095 \\
		w/o raw input & 30.05 & 0.7827 & 30.51 & 0.7921  \\
		w/o feature fusion & {\color{blue}30.73} & {\color{blue}0.8025} & {\color{blue}31.68} & {\color{blue}0.8252}  \\
		Ours full model & {\color{red}30.79} & {\color{red}0.8044} & {\color{red}31.79} & {\color{red}0.8272}  \\
		\hline
	\end{tabular}
\end{table}

\subsection{Results on Synthetic Data}
We quantitatively evaluate our method using the test dataset described above.
Table~\ref{tab:syn1} shows that the proposed algorithm performs favorably against the baseline methods in terms of both PSNR and structural similarity (SSIM). 
Figures~\ref{fig:synthetic} shows some restored results from our synthetic dataset. 
Since the low-resolution raw input does not have the relevant information for color conversion and tone adjustment conducted within the ISP system, the raw image processing methods~\cite{learning_to_see_in_the_dark,deepisp_tip} can only approximate the color corrections of one specific camera, and thereby cannot recover good results for different camera types in our test set. 
In addition, the training process of the raw processing models is influenced by the color differences between the predictions and ground truth, and cannot focus on restoring sharp structures. Thus, the results in Figures~\ref{fig:synthetic}(c) are still blurry.
Figure~\ref{fig:synthetic}(d)-(e) show that the super-resolution methods with low-resolution color image as input generate shaper results with correct colors, which, however, still lack fine details.
In contrast, our method achieves better results with clear structures and fine details in Figure~\ref{fig:synthetic}(f) by exploiting both the color input and raw radiance information.

\begin{figure*}[t]
	\footnotesize
	\begin{center}
		\begin{tabular}{c}
			\includegraphics[width = 0.99\linewidth]{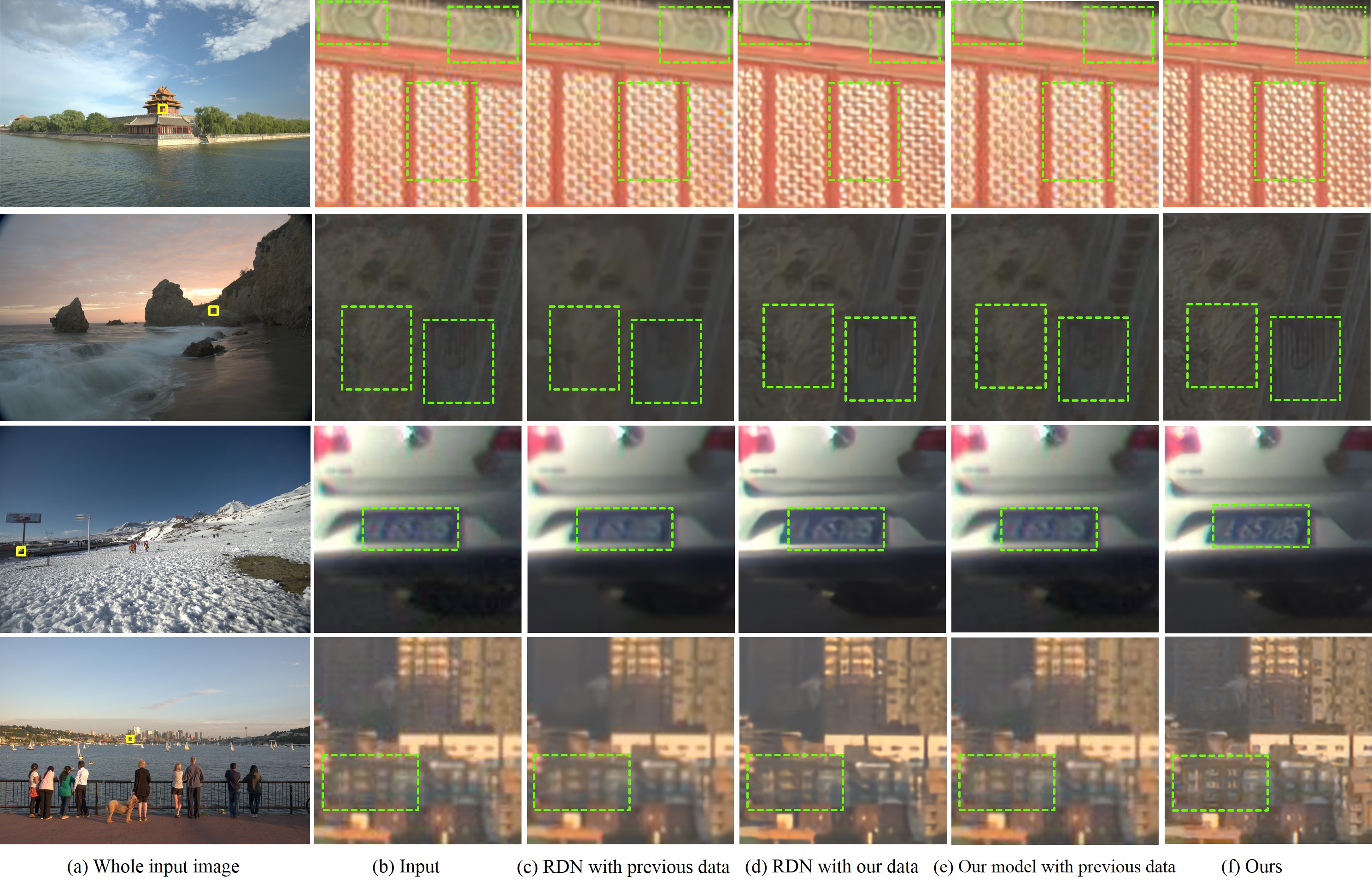} \\
		\end{tabular}
	\end{center}
	\caption{%
		Results on real images. Since the outputs are of ultra-high resolution, typically $8000\times12000$ for a Leica camera and $4000\times 6000$ for Canon, we only show image patches cropped from the tiny yellow boxes in (a). The images from top to bottom are captured by Canon, Sony, Nikon and Leica cameras, respectively.
	}
	\label{fig:real}
\end{figure*}

\vspace{1ex}
{\noindent \bf{Non-blind super-resolution.}} %
Since most super-resolution methods~\cite{SRCNN,SR-VDSR,SR-residual-dense} are non-blind with the assumption that the downsampling blur kernel is known and fixed,
we also evaluate our method for this case by using fixed defocus kernel with radius 5 to synthesize training and test datasets.
As shown in Table~\ref{tab:syn1}, the proposed method achieves competitive super-resolution results under non-blind settings.

\vspace{1ex}
{\noindent \bf{Learning more complex color corrections.}} %
In addition to the reference images rendered by Dcraw,
we also train the proposed model on manually retouched images by photography artists from the MIT-Adobe 5K dataset~\cite{adobe-mit-5k},  
which represent more diverse ISP systems.
We test our model on the same raw input with different reference images produced by different artists in Figure~\ref{fig:artistic}.
The proposed algorithm generates clear structures as well as high-fidelity colors, which shows that our method is able to generalize to more diverse and complex ISP systems.

\subsection{Ablation Study}
For better evaluation of the proposed algorithm,
we test variant versions of our model by removing each component.
We show the quantitative results on the synthetic datasets in Table~\ref{tab:syn1} and provide qualitative comparisons in Figure~\ref{fig:ablation}.
First, without the color correction branch, the network directly uses the low-resolution raw input for super-resolution, which cannot effectively recover high-fidelity colors for diverse cameras (Figure~\ref{fig:ablation}(c)).
Second, simply adopting the global color correction strategy from~\cite{deepisp_tip} could only recover holistic color appearances, such as in the background of Figure~\ref{fig:ablation}(d), but there are still significant color errors in local regions without the proposed pixel-wise transformations.
Third, to evaluate the importance of the raw input, we use the low-resolution color image as the input for both branches of our network, which is enabled by adaptively changing the packing layer of the image restoration branch.
As shown in Figure~\ref{fig:ablation}(e), the model without raw input cannot generate fine structures and details due to the information loss in the ISP system.
In addition, the network without feature fusion cannot predict accurate transformations and tends to bring color artifacts around subtle structures in Figure~\ref{fig:ablation}(f). 
By contrast, our full model effectively integrates different components, and generates sharper results with better details and fewer artifacts in Figure~\ref{fig:ablation}(g) by exploiting the complementary information in raw data and preprocessed color images.

\subsection{Effectiveness on Real Images}
We qualitatively evaluate our data generation method as well as the proposed network on real captured images.
As shown in Figure~\ref{fig:teaser}(d) and \ref{fig:real}(d),
the results of RDN become sharper after re-training with the data generated by our pipeline.
On the other hand, 
the proposed dual CNN model cannot generate clear images (Figure~\ref{fig:real}(e)) by training on previous data generated by bicubic downsampling, homoscedastic Gaussian noise and non-linear space mosaicing~\cite{SRCNN,demosaicking_denoising_deep}.
By contrast, we achieve better results with sharper edges and finer details by training the proposed network on our synthetic dataset as shown in Figure~\ref{fig:teaser}(e) and \ref{fig:real}(f), which demonstrates the effectiveness of the proposed data generation pipeline as well as the raw super-resolution algorithm.
Note that the real images are captured by different types of cameras, and all of them are not seen during training.

\section{Conclusions}
We propose a new pipeline to generate realistic training data by simulating the imaging process of digital cameras.
In addition, we develop a dual CNN to exploit the radiance information recorded in raw data.
The proposed algorithm compares favorably against state-of-the-art  methods both quantitatively and qualitatively, and more importantly, enables super-resolution for real captured images.
This work shows the superiority of learning with raw data, and we expect more applications of our work in other image processing problems.

{\small
\bibliographystyle{ieee}
\bibliography{egbib}

\begin{thebibliography}{10}\itemsep=-1pt

\bibitem{adobe-mit-5k}
V.~Bychkovsky, S.~Paris, E.~Chan, and F.~Durand.
\newblock Learning photographic global tonal adjustment with a database of
  input/output image pairs.
\newblock In {\em CVPR}, 2011.

\bibitem{learning_to_see_in_the_dark}
C.~Chen, Q.~Chen, J.~Xu, and V.~Koltun.
\newblock Learning to see in the dark.
\newblock In {\em CVPR}, 2018.

\bibitem{dcraw}
D.~Coffin.
\newblock Dcraw: Decoding raw digital photos in linux.
\newblock http://www.cybercom.net/~dcoffin/dcraw/.

\bibitem{SRCNN}
C.~Dong, C.~C. Loy, K.~He, and X.~Tang.
\newblock Learning a deep convolutional network for image super-resolution.
\newblock In {\em ECCV}, 2014.

\bibitem{SRCNN_pami}
C.~Dong, C.~C. Loy, K.~He, and X.~Tang.
\newblock Image super-resolution using deep convolutional networks.
\newblock {\em TPAMI}, 38:295--307, 2016.

\bibitem{FSRCNN}
C.~Dong, C.~C. Loy, and X.~Tang.
\newblock Accelerating the super-resolution convolutional neural network.
\newblock In {\em ECCV}, 2016.

\bibitem{farsiu2006multiframe}
S.~Farsiu, M.~Elad, and P.~Milanfar.
\newblock Multiframe demosaicing and super-resolution of color images.
\newblock {\em TIP}, 15:141--159, 2006.

\bibitem{xxy-eccv18-monodepth}
Y.~Gan, X.~Xu, W.~Sun, and L.~Lin.
\newblock Monocular depth estimation with affinity, vertical pooling, and label
  enhancement.
\newblock In {\em ECCV}, 2018.

\bibitem{demosaicking_denoising_deep}
M.~Gharbi, G.~Chaurasia, S.~Paris, and F.~Durand.
\newblock Deep joint demosaicking and denoising.
\newblock {\em ACM Transactions on Graphics (TOG)}, 35:191, 2016.

\bibitem{glorot2010understanding}
X.~Glorot and Y.~Bengio.
\newblock Understanding the difficulty of training deep feedforward neural
  networks.
\newblock In {\em AISTATS}, 2010.

\bibitem{geo_optics}
J.~E. Greivenkamp.
\newblock {\em Field guide to geometrical optics}, volume~1.
\newblock SPIE Press Bellingham, WA, 2004.

\bibitem{hdr+}
S.~W. Hasinoff, D.~Sharlet, R.~Geiss, A.~Adams, J.~T. Barron, F.~Kainz,
  J.~Chen, and M.~Levoy.
\newblock Burst photography for high dynamic range and low-light imaging on
  mobile cameras.
\newblock {\em ACM Trans. Graph. (SIGGRAPH Asia)}, 35(6), 2016.

\bibitem{healey1994radiometric}
G.~E. Healey and R.~Kondepudy.
\newblock Radiometric ccd camera calibration and noise estimation.
\newblock {\em TIP}, 16:267--276, 1994.

\bibitem{AHD2005}
K.~Hirakawa and T.~W. Parks.
\newblock Adaptive homogeneity-directed demosaicing algorithm.
\newblock {\em TIP}, 14:360--369, 2005.

\bibitem{densenet}
G.~Huang, Z.~Liu, L.~Van Der~Maaten, and K.~Q. Weinberger.
\newblock Densely connected convolutional networks.
\newblock In {\em CVPR}, 2017.

\bibitem{jiang2017learning}
H.~Jiang, Q.~Tian, J.~Farrell, and B.~A. Wandell.
\newblock Learning the image processing pipeline.
\newblock {\em TIP}, 26:5032--5042, 2017.

\bibitem{demosaic_denoise_randomfields}
D.~Khashabi, S.~Nowozin, J.~Jancsary, and A.~W. Fitzgibbon.
\newblock Joint demosaicing and denoising via learned nonparametric random
  fields.
\newblock {\em TIP}, 23:4968--4981, 2014.

\bibitem{SR-VDSR}
J.~Kim, J.~K. Lee, and K.~M. Lee.
\newblock Accurate image super-resolution using very deep convolutional
  networks.
\newblock In {\em CVPR}, 2016.

\bibitem{kingma2014adam}
D.~Kingma and J.~Ba.
\newblock Adam: A method for stochastic optimization.
\newblock {\em CoRR}, abs/1412.6980, 2014.

\bibitem{SR-laplacian}
W.-S. Lai, J.-B. Huang, N.~Ahuja, and M.-H. Yang.
\newblock Deep laplacian pyramid networks for fast and accurate
  superresolution.
\newblock In {\em CVPR}, 2017.

\bibitem{xxy-icip18-temporal}
C.~Liu, X.~Xu, and Y.~Zhang.
\newblock Temporal attention network for action proposal.
\newblock In {\em ICIP}, 2018.

\bibitem{leakyRelu}
A.~L. Maas, A.~Y. Hannun, and A.~Y. Ng.
\newblock Rectifier nonlinearities improve neural network acoustic models.
\newblock In {\em ICML}, 2013.

\bibitem{KPN}
B.~Mildenhall, J.~T. Barron, J.~Chen, D.~Sharlet, R.~Ng, and R.~Carroll.
\newblock Burst denoising with kernel prediction networks.
\newblock In {\em CVPR}, 2018.

\bibitem{raw-reconstruction}
R.~M. Nguyen and M.~S. Brown.
\newblock Raw image reconstruction using a self-contained srgb-jpeg image with
  only 64 kb overhead.
\newblock In {\em CVPR}, 2016.

\bibitem{pennebaker1992jpeg}
W.~B. Pennebaker and J.~L. Mitchell.
\newblock {\em JPEG: Still image data compression standard}.
\newblock Springer Science \& Business Media, 1992.

\bibitem{benchmarking_denoising}
T.~Plotz and S.~Roth.
\newblock Benchmarking denoising algorithms with real photographs.
\newblock In {\em CVPR}, 2017.

\bibitem{xxy-tip19-dehazing}
W.~Ren, J.~Zhang, X.~Xu, L.~Ma, X.~Cao, G.~Meng, and W.~Liu.
\newblock Deep video dehazing with semantic segmentation.
\newblock {\em TIP}, 28:1895--1908, 2019.

\bibitem{schuler2013machine}
C.~Schuler, H.~Burger, S.~Harmeling, and B.~Scholkopf.
\newblock A machine learning approach for non-blind image deconvolution.
\newblock In {\em CVPR}, 2013.

\bibitem{deepisp_tip}
E.~Schwartz, R.~Giryes, and A.~M. Bronstein.
\newblock Deepisp: Towards learning an end-to-end image processing pipeline.
\newblock {\em TIP}, 2018.

\bibitem{shi2016real}
W.~Shi, J.~Caballero, F.~Husz{\'a}r, J.~Totz, A.~P. Aitken, R.~Bishop,
  D.~Rueckert, and Z.~Wang.
\newblock Real-time single image and video super-resolution using an efficient
  sub-pixel convolutional neural network.
\newblock In {\em CVPR}, 2016.

\bibitem{SR-regression-accv14}
R.~Timofte, V.~De~Smet, and L.~Van~Gool.
\newblock A+: Adjusted anchored neighborhood regression for fast
  super-resolution.
\newblock In {\em ACCV}, 2014.

\bibitem{SR-dense}
T.~Tong, G.~Li, X.~Liu, and Q.~Gao.
\newblock Image super-resolution using dense skip connections.
\newblock In {\em ICCV}, 2017.

\bibitem{vandewalle2007joint}
P.~Vandewalle, K.~Krichane, D.~Alleysson, and S.~S{\"u}sstrunk.
\newblock Joint demosaicing and super-resolution imaging from a set of
  unregistered aliased images.
\newblock In {\em Digital Photography III}, 2007.

\bibitem{SR-SCN}
Z.~Wang, D.~Liu, J.~Yang, W.~Han, and T.~Huang.
\newblock Deep networks for image super-resolution with sparse prior.
\newblock In {\em ICCV}, 2015.

\bibitem{xxy-tip18-edgedeblur}
X.~Xu, J.~Pan, Y.-J. Zhang, and M.-H. Yang.
\newblock Motion blur kernel estimation via deep learning.
\newblock {\em TIP}, 27:194--205, 2018.

\bibitem{xxy-eccv18-rendering}
X.~Xu, D.~Sun, S.~Liu, W.~Ren, Y.-J. Zhang, M.-H. Yang, and J.~Sun.
\newblock Rendering portraitures from monocular camera and beyond.
\newblock In {\em ECCV}, 2018.

\bibitem{xxy-iccv17}
X.~Xu, D.~Sun, J.~Pan, Y.~Zhang, H.~Pfister, and M.-H. Yang.
\newblock Learning to super-resolve blurry face and text images.
\newblock In {\em ICCV}, 2017.

\bibitem{SR-residual-dense}
Y.~Zhang, Y.~Tian, Y.~Kong, B.~Zhong, and Y.~Fu.
\newblock Residual dense network for image super-resolution.
\newblock In {\em CVPR}, 2018.

\bibitem{SR_demosaic_deep}
R.~Zhou, R.~Achanta, and S.~S{\"u}sstrunk.
\newblock Deep residual network for joint demosaicing and super-resolution.
\newblock {\em arXiv preprint arXiv:1802.06573}, 2018.

\end{thebibliography}
}

\end{document}